\documentclass[a4paper,11pt]{article}

\usepackage[latin1]{inputenc}
\usepackage[T1]{fontenc}
\usepackage{textcomp,graphicx,epsfig}
\usepackage{latexsym,amssymb,amsmath,amsthm}
\usepackage{fancyhdr,rotating,makeidx,eurosym}
\usepackage{lmodern,url,xspace,hyperref,layout}

\def\PP{\mathbb{P}}

\def\dd{\mathrm{d}}
\def\ee{\mathrm{e}}

\newenvironment{algo}{ 
                      \vskip 2mm \noindent \hspace{-2mm}\sf
                      }
                     { 
                     \vskip 2mm \noindent   
                     }

\newcommand{\tab}{\hspace*{1cm}} 

\begin{document}

\title{Simulation of Gene Regulatory Networks}
\author{Bernard Ycart\footnote{Laboratoire Jean Kuntzmann, 
Univ. Grenoble-Alpes, France \texttt{Bernard.Ycart@imag.fr}}%
       \and 
Fr\'ed\'eric Pont\footnote{INSERM UMR1037-Cancer 
Research Center of Toulouse, France%
         \texttt{Frederic.Pont@inserm.fr}}%
       \and 
Jean-Jacques Fourni\'e\footnote{INSERM UMR1037-Cancer 
Research Center of Toulouse, France%
         \texttt{Jean-Jacques.Fournie@inserm.fr}}}%

\date{\vspace{-5ex}}

\maketitle


\begin{abstract}
This limited review is intended as an introduction 
to the fast growing subject of mathematical modelling of cell
metabolism and its biochemical pathways, and more precisely on
pathways linked to apoptosis of cancerous cells. 
Some basic mathematical models of chemical kinetics, with emphasis
on stochastic models, are presented.
\end{abstract}

\noindent
\textit{Keywords:} biochemical pathways; kinetics; master equation;
  diffusion processes; simulation

\noindent
\textit{MSC:} 80A30; 92C40
\section{Introduction}
This is intented as an introduction to the rapidly growing literature
on the mathematical modelling of biochemical pathways. This subject links
several, quite different areas of research. Depending on viewpoints, relevant
articles can be found in Biology, Chemistry,
Mathematics, or Informatics journals. Rather than seeking a hopeless
exhaustivity, we have tried to illustrate some of the current 
approaches by a few recent references.

Inside the huge domain of biochemistry, we have focused on cell
metabolism and its biochemical pathways, and more precisely on
pathways linked to apoptosis of cancerous cells. On the other hand,
mathematical modelling in genomics
has used many different techniques,
among which we chose to restrict our study to dynamic equations,
insisting on stochastic models.

The paper is organized as follows. In the next section we shall review
a few articles on biochemical pathways, enlighting the use of modelling
and in silico experiments. Section 3 presents
the basic mathematical models of chemical kinetics, with emphasis
on stochastic models.
Section 4 deals with  the computer simulation of stochastic models,
with emphasis on Gillespie's algorithm. Computer integration aspects
are treated in section 5. 
\section{Biochemical pathways}
In the ever growing literature on cell metabolism and biochemical pathways,
we have selected a few references dealing with the targeting of
regulatory pathways, in particular apoptosis pathways, 
and their mathematical modelling. 
The global human metabolic map has been reconstructed accounting for
the functions of 1496 ORF's, 2004 proteins, 2766 metabolites, and 3311
metabolic and transport reactions \cite{Duarteetal07}.
A good review on apoptosis pathways is given 
 by Elmore \cite{Elmore07}.

Among the most recent advances, Folger \emph{et al.} \cite{Folgeretal11}
announce a genome-scaled network model of cancer metabolism that they
validate by predicting 52 cytostatic drug targets. 
Shaughnessy \emph{et al.} \cite{Shaughnessy11} describe a way of
artificially regulating MAP kinase cascades.

Clarke \emph{et al.} \cite{Clarkeetal08} use statistical model checking
for analysing t-cell receptor signalling pathways.

Studies linking biochemical pathways to cancer outcomes include 
that of Chuang \emph{et al.} \cite{Chuangetal07} and
Taylor \emph{et al.} \cite{Tayloretal09}.

Devun \emph{et al.} \cite{Devunetal10} have recently
identified a ``mitochondrial permeability transition pore-dependent 
pathway to selective cell death''. 
Teams who have used apoptosis pathways for cancer therapy include
Chu and Chen \cite{ChuChen08},
Ghobrial \emph{et al.} \cite{Ghobrialetal05},
Speirs \emph{et al.} \cite{Speirsetal11}. A success obtained by
combining
Haem oxygenase with fumarate hydratase has recently been announced by 
Frezza \emph{et al.} \cite{Frezzaetal11}.
Yosef \emph{et al.} \cite{Yosefetal09} address the problem of
optimizing gene networks, with application to apoptosis.

The ultimate goal to systematically predict through mathematical 
modelling, possible gene targets that could induce cancerous cell 
apoptosis has been pursued by several teams among which 
Legewie \emph{et al.} \cite{Legewieetal06},
Ryu \emph{et al.} \cite{Ryuetal09},
Huber \emph{et al.} \cite{Huberetal11}.
\section{Mathematical models}
Many mathematical models have been applied to genomics. The interested
reader is referred to Shmulevitch and Dougherty's book
\cite{ShmulevitchDougherty07} or the reviews by 
de Jong \cite{deJong02}, Goutsias and Lee \cite{GoutsiasLee07},
Karlebach and Shamir \cite{KarlebachShamir08}. 
Here we will focus exclusively onto the most classical
dynamic models which are those of chemical kinetics. The mathematical
developments date back to the first half of the  twentieth century
(see the historical section of \cite{McQuarrie67}), and
are presented in many textbooks, in particular those of
van Kampen \cite{vanKampen81}, Ethier and Kurtz \cite{EthierKurtz05},
Gardiner \cite{Gardiner04}, Wilkinson \cite{Wilkinson06},
or Kolokoltsov \cite{Kolokoltsov10}. 
A short and clear introduction has been written by Higham
\cite{Higham08}, from which we shall borrow the terminology regarding the main
three ``chemical equations''. As examples, we shall use the
simplest historic models of  Yule  \cite{Yule25} and 
Michaelis-Menten  \cite{MentenMichaelis13,JohnsonGoody11}.
\vskip 2mm
At the basis of the theory, modelling assumptions are made to ensure
that the only quantities of interest be the number of molecules of each
type of reactant simultaneously present: constant volume, well
stirred medium, space homogeneity. As remarked by Higham
\cite{Higham08} section 9.3, these hypotheses could be
questioned when applied in the biological context of the living
cell. Nevertheless, they are usually considered as
unavoidable, and the models have been theoretically established on a
 very firm physical and chemical base: see for instance Gillespie
 \cite{Gillespie92}. 
 \vskip 2mm
A chemical reaction system is made of a certain number $n$ of
\emph{reactants} (or chemical species), and
another number $m$ of \emph{reactions} destroying or 
producing molecules of them. The $n$-dimensional 
\emph{state vector} $X(t)$ keeps
track of the quantities of each species simultaneously present at
time $t$: its $i$-th entry counts how many molecules of species $i$
are in the solution at time $t$. According to the scale of time and
numbers, three different models are considered. 
\begin{enumerate}
\item \emph{Microscopic scale:} the vector $X(t)$ is a stochastic
  process with birth-and-death dynamics: reactions take place
  one at a time at random instants and modify $X(t)$ by a few units added or
  substracted from some coordinates. 
The probabilities of all possible states are the
  solution of the Chapmann-Kolmogorov equations for the Markov process
  $\{X(t)\}$. That system of Ordinary Differential Equations (ODE's)
  is called the \emph{Chemical Master Equation}.
\item \emph{Mesoscopic scale:} when rescaling time and space, the Markov
  jump process $\{X(t)\}$ converges to a continuous real-valued
  diffusion process, solution of 
  a system of Stochastic Differential Equations (SDE's), called the
  \emph{Chemical Langevin Equation}.
\item \emph{Macroscopic scale:} when stochastic fluctuations are
  neglected, the quantities of molecu\-les are viewed as derivable
  functions of time, solution of a system of ODE's, called the
  \emph{Reaction Rate Equations}.
\end{enumerate}
The Chemical Master Equation can be explicitly solved
  only in exceedingly simple, mostly irrelevant cases, such as
  the Yule model that we shall treat below (other examples are
  reviewed by McQuarrie in section III of \cite{McQuarrie67}). 
  The alternative approach is to
  simulate trajectories of $X(t)$ through Gillespie's algorithm,
  that we shall examine in more detail in the next section. As pointed
  out in section 6 of Higham \cite{Higham08}, the so called
  ``tau-leaping'' method, which is the usual way of
  accelerating the algorithm, can be viewed as an approximate
  solution to the Chemical Langevin Equation. On the theoretical side,
convergence of birth-and-death dynamics to diffusion processes has been
proved by Kurtz at the end of the 70's: chapter 10 of
  \cite{EthierKurtz05} gives a complete theoretical treatment of
diffusion processes in chemical kinetics. Thus, from the modelling point
of view, it can be considered that the microscopic and mesoscopic
scales are not essentially different: discrepancies come
from the mathematical  or algorithmic treatment. 
The real opposition is between stochastic (microscopic and mesoscopic
scales) and deterministic (macroscopic scale) modelling. It has been
discussed at length in  many references, see for instance Chen \emph{et
  al.} \cite{Chenetal10}, Goutsias \cite{Goutsias07}, Liang and Qian
\cite{LiangQian10}, Qian \cite{Qian00} or
Shahrezaei and Swain \cite{Shahrezaei08}. The main reason why
stochastic is usually preferred to deterministic for
gene regulatory networks lies in the order of magnitude of
the numbers of molecules involved. The proteins produced by mRNA can
usually be counted by a few hundreds per second, far short from Avogadro's
number ($6.02~10^{23}$) which is the scale at which deterministic
models work.
\vskip 2mm
We shall now introduce the main concepts of mathematical kinetics models,
illustrating them on the two basic examples of the Yule and
Michaelis-Menten models. 
The effect of reactions on species is usually described by symbolic
equations of the type $A+B\longrightarrow C$, meaning that
each time the reaction takes place or \emph{fires}, one molecule
of $A$ and one molecule of $B$ combine to form one molecule of $C$. For
mathematical purposes, it is convenient to encode the effect of each
reaction by two vectors of integers, called the 
\emph{stoichiometric
vectors} (from the Greek meaning ``measure of
elements''). To the $j$-th reaction correspond vectors $\nu^-_j$ and
$\nu^+_j$. Their entries are as follows.
\begin{itemize}
\item
$\nu^-_j(i)$ is the number of molecules of species $i$ that
are \emph{destroyed} by reaction $j$,
\item
$\nu^+_j(i)$ is the number of molecules of species $i$ that
are \emph{produced} by reaction $j$.
\end{itemize}
For instance reaction $A+B\longrightarrow C$ will be translated by
two 3-dimensional vectors, with entries indexed by $A$ $B$ and $C$.
$$
\nu^- = \left(\begin{array}{c}1\\1\\0\end{array}\right)
\quad\mbox{and}\quad
\nu^+ = \left(\begin{array}{c}0\\0\\1\end{array}\right)
$$
\vskip 2mm
The \emph{Yule model} is a very basic population model used for instance in
 the case of bacteria population growth. Formally, it could be assimilated to a
 chemical equation with only one
 reactant: $A\longrightarrow 2A$ (one bacteria
 becomes two when the reaction/meiosis fires). The stoichiometric
 vectors only have one entry: $\nu^-=1$ and $\nu^+=2$.
\vskip 2mm
The \emph{Michaelis-Menten} equations involve four species:
\begin{itemize}
\item a substrate $S$
\item an enzyme $E$
\item a complex $C$
\item a product $P$
\end{itemize}

Here are the three equations and the corresponding stoichiometric vectors,
indexed by\\ $S,E,C,P$ in that order.
\begin{enumerate}
\item $S+E\longrightarrow C$:
$\qquad\nu^-_1=\left(\begin{array}{c}1\\1\\0\\0\end{array}\right)$,
$\nu^+_1=\left(\begin{array}{c}0\\0\\1\\0\end{array}\right)$.
\item $C\longrightarrow S+E$,
$\qquad\nu^-_2=\left(\begin{array}{c}0\\0\\1\\0\end{array}\right)$,
$\nu^+_2=\left(\begin{array}{c}1\\1\\0\\0\end{array}\right)$.
\item $C\longrightarrow E+P$,
$\qquad\nu^-_3=\left(\begin{array}{c}0\\0\\1\\0\end{array}\right)$,
$\nu^+_3=\left(\begin{array}{c}0\\1\\0\\1\end{array}\right)$.
\end{enumerate}
Some textbooks use the more compact \emph{stoichiometric matrix}
that summarizes in its columns the vectors above. Here is that matrix
for the Michaelis-Menten model.
$$
\begin{array}{rrr|r}
1&2&3&\\
\hline
-1&+1&0&S\\
-1&+1&+1&E\\
+1&-1&-1&C\\
0&0&+1&P
\end{array}
$$
There are two reasons to prefer the vectors $\nu^+$
and $\nu^-$. One is that the matrix looses information when the same
species appears on both sides of the equation (as in the Yule
model). The other is that the propensities, to be
defined later, depend only on the $\nu^-_j$'s and not on the
$\nu^+_j$'s.
\vskip 2mm
Regarding the mathematical expression of the different models, 
we shall follow the
introduction of Ball \emph{et al.} \cite{Balletal06}. For the
stochastic version of Michaelis-Menten dynamics, see Qian
\cite{Qian02}, Sanft \emph{et al.} \cite{Sanft11}
and Higham \cite{Higham08}.
Once the stoichiometry of the system is known, the evolution of the
state vector only depends on successive reaction firings. Denote by
$R_j(t)$ the number of times reaction $j$ fires between $0$ and
$t$. Then:
$$
X(t) = X(0)+\sum_{j=1}^m R_j(t)(\nu_j^+-\nu_j^-)\;.
$$ 
The $\{R_j(t)\}$'s are counting processes, whose instantaneous rates
are called the \emph{propensities}. Intuitively, during an interval of
time $[t,t+\delta t]$ short enough to ensure that only one reaction
will fire in that interval, the probability that reaction $j$
fires should be proportional to the duration $\delta t$. The
proportionality coefficient is the instantaneous rate of
$R_j(t)$. Assuming the system is well stirred, all molecules are
equally likely to be at any location at any time. So the probability
that all molecules required for reaction $j$ meet, should be
proportional to the number of ways of finding these molecules
together. Let $x$ be the vector of integers giving the number of
molecules of each species. The propensity of reaction $j$ when the
state vector is $x$ is:
$$
\lambda_j(x) = \kappa_j\,\frac{N}{N^{\sum_i\nu^-_j(i)}} \,\prod_i\nu^-_j(i)!
\binom{x}{\nu^-_j(1)\ldots\nu^-_j(n)}\;,
$$
where $N$ is a scaling parameter, usually taken to be the volume of
the system multiplied by Avogadro's number, and $\kappa_j$ is a constant, 
depending only on the reaction (including of course temperature 
conditions). 
Rather than detailing
the general formula above, we shall make its meaning clear on simple
examples, involving only 2 species $A$ and $B$, for which the number
of molecules are denoted by $a$ and $b$. 
\begin{itemize}
\item 
$A\,\longrightarrow\,\cdots~$: $\lambda_j(x)=\kappa_j\, a$  
\item 
$2A\,\longrightarrow\,\cdots~$: $\lambda_j(x)=\kappa_j\,\frac{1}{N}\,a(a-1)$
\item 
$A+B\,\longrightarrow\,\cdots~$: $\lambda_j(x)=\kappa_j\,\frac{1}{N}\, ab$  
\item 
$2A+3B\,\longrightarrow\,\cdots~$: $\lambda_j(x)=\kappa_j\,\frac{1}{N^4}\,a(a-1)
\,b(b-1)(b-2)$  
\end{itemize}
Using propensities, the counting processes $\{R_j(t)\}$ can be written
as:
$$
R_j(t) = Y_j\left(\int_0^t \lambda_j(X(s))\,\dd s\right)\;,
$$
where the $Y_j$ are independent unit Poisson processes. Hence the
\emph{Stochastic Integral Equation} (SIE) defining $X$:
\begin{equation*}
\tag{SIE}
X(t) = X(0)+ \sum_{j=1}^m 
Y_j\left(\int_0^t \lambda_j(X(s))\,\dd s\right)(\nu_j^+-\nu_j^-)\;.
\end{equation*}
The \emph{Chemical Master Equation} (CME) is the forward 
Chapmann-Kolmogorov equation
corresponding to the Markov process $\{ X(t)\}$. It 
is a linear system of ODE's where the
unknowns are the probabilities for $X(t)$ to be
in each possible state and the propensities are the coefficients.
\begin{equation*}
\tag{CME}
\begin{array}{lcl}
\displaystyle{\frac{\dd}{\dd t}\PP[X(t)=x]}
&=&\displaystyle{-\left(\sum_j \lambda_j(x)\right)\PP[X(t)=x]}\\
&&\displaystyle{+ \sum_j \lambda_j(x-\nu^+_j+\nu^-_j)\, 
\PP[X(t)=x-\nu^+_j+\nu^-_j]\;.}
\end{array}
\end{equation*}
Here is the CME for the Yule model $A\longrightarrow 2A$, 
denoting by $p_n(t)$ the probability that $n$ molecules of $A$ are
present at time $t$.
$$
\frac{\dd p_n(t)}{\dd t} = -\kappa n p_n(t) +\kappa(n-1)p_{n-1}(t)\;.
$$
It is a well know fact that this equation admits an
explicit solution: given an initial number of molecules
of $n_0$, the number of molecules at time $t$ follows the
negative binomial distribution with parameters $n_0$ and 
$\ee^{-\kappa t}$:
$$
p_n(t) = \binom{n-1}{n_0-1} \ee^{-\kappa  n_0 t}(1-\ee^{-\kappa  t})^{n-n_0}\;.
$$ 
The expectation of that distribution is $n_0\ee^{\kappa t}$, and its
variance is $n_0(1-\ee^{-\kappa t})\ee^{2\kappa t}$.
As already mentioned, cases like this one are very rare and no explicit
solution to the CME exists in general.
\vskip 2mm\noindent
We shall now introduce the renormalization that leads to the Chemical
Langevin Equation. Let $x$ be a value of the state vector (counting
molecules of each species). If $N$ is the volume multiplied by
 Avogadro's number, then
$
c = \frac{1}{N}x
$
is the vector of \emph{concentrations} (expressed in moles per unit
volume). If the coordinates of $x$ are large, then the propensities
can be expressed asymptotically as follows:
$$
\lambda_j(x) \simeq N \kappa_j \prod_i c(i)^{\nu^-_j(i)}\;.
$$ 
Hence we can define ``macroscopic'' propensities as:
$$
\widetilde{\lambda}_j(c) = \kappa_j  \prod_i c(i)^{\nu^-_j(i)}\;.
$$
Recall the SIE defining $X$, and divide both members by $N$:
$$
\frac{X(t)}{N} = \frac{X(0)}{N}+ \sum_{j=1}^m 
\frac{1}{N}Y_j\left(\int_0^t \lambda_j(X(s))\,\dd s\right)(\nu_j^+-\nu_j^-)\;.
$$ 
Replacing $X$ by the concentration vector $C=\frac{1}{N}X$ gives:
$$
C(t) = C(0)+ \sum_{j=1}^m 
\frac{1}{N}Y_j\left(\int_0^t N\widetilde{\lambda}_j(C(s))\,\dd
  s\right)
(\nu_j^+-\nu_j^-)\;.
$$ 
Now the central limit theorem describes the asymptotics for the unit 
Poisson process (random counting with an average count of $1$ per time
unit). At (large) time $Nu$ the process reaches $Nu$ on average with
fluctuations of order $\sqrt{N}$ described by a standard Brownian motion.  
$$
\lim_{n\to +\infty} \frac{Y(Nu)-Nu}{\sqrt{N}} = W(u)\;,
$$
where $Y$ is a unit Poisson process, $W$ is the standard Brownian
motion and the limit is understood in distribution. So for each
reaction $j$, we can replace $Y_j(Nu)$ by $Nu+\sqrt{N}W(u)$ and
get the approximation:
$$
\frac{1}{N}Y_j\left(\int_0^t N\widetilde{\lambda}_j(C(s))\,\dd
  s\right) \simeq 
 \int_0^t \widetilde{\lambda}_j(C(s))\,\dd s
+
\frac{1}{\sqrt{N}}W_j\left(\int_0^t 
\widetilde{\lambda}_j(C(s))\,\dd s\right)\;.
$$
The definition of $C(t)$ at mesoscopic scale will thus be:
\begin{eqnarray*}
C(t) &=& \displaystyle{C(0) + 
\left(\int_0^t \sum_j \widetilde{\lambda}_j(C(s))
(\nu^+_j-\nu^-_j)\,\dd s\right)}\\[1.5ex]
&&\displaystyle{+
\frac{1}{\sqrt{N}} 
\left(\sum_jW_j\left(\int_0^t \widetilde{\lambda}_j(C(s))\,\dd s\right)
(\nu^+_j-\nu^-_j)\,\right)\;.}
\end{eqnarray*}
This diffusion process is a solution to the following SDE, called
\emph{Chemical Langevin Equation} (CLE).
\begin{equation*}
 \tag{CLE}
\begin{array}{lcl}
\dd C(t) &=&  \displaystyle{
\left(\sum_j \widetilde{\lambda}_j(C(s))
(\nu^+_j-\nu^-_j)\right)\,\dd t}\\[1.5ex]
&&\displaystyle{+
\frac{1}{\sqrt{N}} 
\left(\sum_j \sqrt{\widetilde{\lambda}_j(C(s))}(\nu^+_j-\nu^-_j)\,\dd W_j
\,\right)\;.}
\end{array}
\end{equation*}
By neglecting the stochastic term in the CLE, we get the classical
\emph{Reaction Rate Equation} (RRE).
\begin{equation*}
\tag{RRE}
\frac{\dd}{\dd t} C(t) = \sum_j \widetilde{\lambda}_j(C(s))
(\nu^+_j-\nu^-_j)\;.
\end{equation*}
When the first member of the RRE vanishes, there remains a system of
algebraic equations, relatively easy to solve, at least
numerically. Its solution (which is a solution to the RRE
\emph{constant in time}) is called a \emph{steady state}, or chemical
equilibrium. Not all systems admit a steady state, but general
conditions of existence have been found: see Feinberg
\cite{Feinberg95}. Stiefenhofer \cite{Stiefenhofer98}, 
Thomson and Gudawardena \cite{Thomson09a,Thomson09b} study a
theoretical approach to the system of algebraic equations involved
in the computation of steady states in the context of biochemical systems.
\vskip 2mm
As an illustration, we give below the CLE of our two examples. For
the Yule model:
$$
\dd a(t) = \kappa a(t) \,\dd t + \sqrt{\kappa a(t)}\, \dd W
$$
Notice that in this case, the RRE $\dd a(t) = \kappa a(t) \dd t$ gives
$a(t)=a(0)\ee^{\kappa t}$, in accordance with the expectation of
$X(t)$ given above. This is a particular case: the expectation of the
distribution solving the CME or the CLE is not the solution of the
RRE in general. The relations between different types of models has
been analyzed by Gillespie \emph{et al.} \cite{Gillespieetal09}.

Here are the macroscopic propensities for the Michaelis-Menten model.
\begin{enumerate}
\item $S+E\longrightarrow C$:
$\widetilde{\lambda}_1 = \kappa_1 s(t)e(t)$,
\item $C\longrightarrow S+E$:
$\widetilde{\lambda}_2 = \kappa_2 c(t)$,
\item $C\longrightarrow E+P$:
$\widetilde{\lambda}_3 = \kappa_3 c(t)$.
\end{enumerate} 
Here is the CLE:
\begin{eqnarray*}
\dd s(t)&=&\displaystyle{
\Big(-\kappa_1 s(t)e(t)+\kappa_2c(t)\Big)\,\dd t}\\
&&\hspace*{2.5cm}\displaystyle{+\frac{1}{\sqrt{N}}
\Big(-\sqrt{\kappa_1 s(t)e(t)}\,\dd W_1+\sqrt{\kappa_2c(t)}\,\dd
W_2\Big)}\\[1.5ex]
\dd e(t)&=&\displaystyle{
\Big(-\kappa_1 s(t)e(t)+\kappa_2c(t)+\kappa_3c(t)\Big)\,\dd t}\\
&&\hspace*{2.5cm}\displaystyle{+\frac{1}{\sqrt{N}}
\Big(-\sqrt{\kappa_1 s(t)e(t)}\,\dd W_1+\sqrt{\kappa_2c(t)}\,\dd W_2
+\sqrt{\kappa_3c(t)}\,\dd W_3\Big)}\\[1.5ex]
\dd c(t)&=&\displaystyle{
\Big(+\kappa_1 s(t)e(t)-\kappa_2c(t)-\kappa_3c(t)\Big)\,\dd t}\\
&&\hspace*{2.5cm}\displaystyle{+\frac{1}{\sqrt{N}}
\Big(\sqrt{\kappa_1 s(t)e(t)}\,\dd W_1-\sqrt{\kappa_2c(t)}\,\dd W_2
-\sqrt{\kappa_3c(t)}\,\dd W_3\Big)}\\[1.5ex]
\dd p(t)&=&\displaystyle{
\Big(\kappa_3c(t)\Big)\,\dd t}\\
&&\hspace*{2.5cm}\displaystyle{+\frac{1}{\sqrt{N}}
\Big(\sqrt{\kappa_3c(t)}\,\dd W_3 \Big)}\;.
\end{eqnarray*}
\vskip 2mm
Even knowing that mathematical models have been established long ago on
very sound theoretical grounds,  one must remain aware
of two major issues. The first one is the combinatorics of reaction
systems involved in gene regulatory networks. The complete cell 
metabolism involves
reactants and reactions by the tens of thousands. No computer program
can solve so large systems of equations, deterministic or stochastic.
In order to find shortcuts that limit the amount of calculations
required, many different approaches have been tempted. Ball \emph{et
  al.} \cite{Balletal06} 
propose to take into account the very different scales of times
at which reactions fire, and distinguish the slower components of a
system from the faster. In the same vein, Goutsias \cite{Goutsias05}
eliminates the effect of the faster reactions by using steady states.  
Grognard \emph{et al.}
\cite{Grognardetal07} replace the essentially non linear
models by piece-wise linear approximations, for which explicit
solutions can be computed. Using the same technique, Ropers \emph{et
  al.} \cite{Ropersetal11} announce interesting stability
results. Among many others, Mallavarapu \emph{et al.} 
\cite{Mallavarapu09} use modularity to divide a full scale model into
more tractable components. This approach consists of considering one
module of interest, while assuming that the rest of the system has
reached a steady state. Gunawardena \cite{Gunawardena11} gives a
graph-theoretical basis to modularity. 
De Jong and Page \cite{deJongPage08} consider steady states in the
context of piece-wise linear approximations.
Rao and Arkin \cite{RaoArkin03} and 
McNamara \emph{et al.} \cite{McNamara08} couple the simulation of 
the CME with steady states approximations. 
Ramaswamy \emph{et al.} \cite{Ramaswamy11b} study the stochastic
fluctuations around the steady state.
\vskip 2mm
The other important issue is the estimation of parameters. The models
presented in the previous section are based on the propensity functions,
parametrized 
by the reaction contants $\kappa_j$. Estimating these constants is not
an easy task. In the slighty different context of viral dynamics, Miao
\emph{et al.} \cite{Miaoetal11} have recently reviewed the different
approaches to that problem, called \emph{identifiability}. 
The most obvious technique, already used by Michaelis and Menten 
(see \cite{JohnsonGoody11}), is regression analysis on experimental data: 
see Jaqaman and 
Danuser \cite{Jaqaman06} for a general review. Batt \emph{et al.}
\cite{Battetal10}  or
de Jong and Ropers \cite{deJongRopers06} propose alternative
approaches to parametric estimation. Even in cases where only raw
estimates of the 
reaction constants were known, some encouraging results show that
approximated simulations can still account at least qualitatively for
experimental results: see Gutenkunst \emph{et al.} \cite{Gutenkunst07}
or Ropers \emph{et al.}
\cite{Ropersetal11}.
\section{Stochastic Simulation Algorithms}
As already pointed out, there exists a structural reason to prefer
stochastic models to deterministic ones for gene regulatory network
modelling. For that reason, and also because numerical solvers for ODE's
are well know and implemented in all mathematical softwares, we shall
not present them here. Solutions to the stochastic models (either the
CME or the CLE) can only be obtained through Monte-Carlo methods,
\emph{i.e.} by simulating the stochastic processes under
consideration. The basic method of simulation is 
\emph{Gillespie's algorithm}, also called 
Stochastic Simulation Algorithm in some
references. Proposed in the context of chemical reactions by
Gillespie \cite{Gillespie76,Gillespie77}, it is a particular case
of the general method for simulating Markov Jump Processes in
continuous time through their imbedded Markov chain. The basic version
simulates a trajectory of the state vector $X(t)$ by translating 
the SIE introduced in the previous section:
\begin{equation*}
\tag{SIE}
X(t) = X(0)+ \sum_{j=1}^m 
Y_j\left(\int_0^t \lambda_j(X(s))\,\dd s\right)(\nu_j^+-\nu_j^-)\;.
\end{equation*}
It should be understood as follows. At any time $s$, all $m$ reactions
have independent firing times, that of reaction $j$ being
exponentially distributed with parameter $\lambda_j(X(s))$. The next
reaction will fire after a time which is the minimum of these
random variables. From elementary properties of exponential random
variables, it can be deduced that: 
\begin{enumerate}
\item the next firing will occur after a random time, exponentially
  distributed with parameter 
$\displaystyle{\lambda_{\bullet}(X(s))=\sum_{j=1}^m \lambda_j(X(s))}$.
\item it will be the firing of reaction $j$ with probability  
$\displaystyle{\frac{\lambda_{j}(X(s))}{\lambda_{\bullet}(X(s))}}$.
\end{enumerate}
The two elementary steps above are easily simulated. Any programming
language has a \textsf{Random} function. Successive calls of that
function output sequences of pseudo-random
numbers that can be considered as realizations of independent random
variables, uniformly distributed on the interval $[0,1]$. Using
\textsf{Random}, the two elementary steps of the simulation are easily
programmed. 
\begin{enumerate}
\item Knowing $\lambda$, to simulate a random variable
  exponentially distributed with parameter $\lambda$, output
$-\log(\mbox{\textsf{Random}})/\lambda$.
\item Knowing $p_1,\ldots,p_m$, to simulate a random index $j$
  with probability $p_j$,
\begin{enumerate}
\item compute cumulated probabilities $q_j=p_1+\cdots+p_j$
\item compare \textsf{Random} to the $q_j$'s
\item if \textsf{Random} is in the interval $[q_{j-1},q_j]$, output $j$.
\end{enumerate}
\end{enumerate}
When reaction $j$ fires, the state vector $X(s)$ is incremented using
the stoichiometric vectors $\nu^-_j$ and $\nu^+_j$.
The pseudo-code of the Gillespie Algorithm can be written as follows.
\begin{algo}
Initialize\\ 
\tab Time scale $T\,\longleftarrow\,0$\\
\tab State vector $X\,\longleftarrow\,X(0)$\\
Repeat\\
\tab Compute propensities $\lambda_j(X)$ for $j=1,\ldots,m$\\
\tab Sum them to get $\lambda_{\bullet}(X)$\\
\tab Compute random delay:
$S\,\longleftarrow\,-\log(\mbox{\textsf{Random}})/\lambda_{\bullet}(X)$\\
\tab Compute probabilities: 
$p_j=\displaystyle{\frac{\lambda_{j}(X(s))}{\lambda_{\bullet}(X(s))}}$\\
\tab Among all reactions, choose reaction $j$ with probability $p_j$\\
\tab Update\\
\tab\tab Time scale: $T\,\longleftarrow\,T+S$\\
\tab\tab State vector: $X\,\longleftarrow\,X+\nu^+_j-\nu^-_j$\\
Until end of simulation
\end{algo}
Higham \cite{Higham08} provides a Matlab implementation of that
algorithm. However, several reasons make it quite inefficient.
\begin{itemize}
\item 
Billions of steps should be simulated to get a trajectory
  comparable to experimental results. Even though each one of these
  steps is relatively cheap, the computer time required for a full
  scale simulation is prohibitive.
\item
Even with an arithmetic coprocessor, the $\log$ function required to
compute the random delay between firings is more expensive than
additions and multiplication. Using
it at each step is time consuming, and essentially useless: from the
central limit theorem, it is known that a sum of independent
random variables is asymptotically distributed as a normal 
variable with same expectation and variance. The expectation of an
exponential with parameter $\lambda$ is $\frac{1}{\lambda}$ and its
variance is $\frac{1}{\lambda^2}$.  
It is much more efficient to cumulate at each step 
$\frac{1}{\lambda}$ and $\frac{1}{\lambda^2}$, then 
update the time scale only at regular
intervals (say after $10^4$ firings), with a normal variable.
\item
If the number of reactions is large (which is the case of gene
regulatory networks), choosing a reaction with a given probability can
be expensive. A lot of computer time is lost in updating probabilities
and cumulating them. Moreover, comparing \textsf{Random} to the
cumulated probabilities results in many useless tests, especially
when the $p_i$'s have very different orders of magnitude. A lot of
computer time can be saved by ranking the $p_i's$ in decreasing
order. But this can be efficient only when the same probability vector is
used many times.
\item
At each step, the state vector is updated by a random vector which is
equal to $\nu^+_j-\nu^-_j$ with probability $p_j$. But again the
central limit theorem can be used. After a large number of steps, the
state vector has been updated by a sum of independent random
variables, the expectation and variance of which can be cumulated to
simulate a normal approximation only at the end of the loop.
\item
Updating the state vector and the propensities at each step is also
time consuming and relatively inefficient, because when the numbers of
molecules are large, changing them by a few units does not essentially
modify the results.   
\end{itemize}
The remarks above where made long ago and lead to the so called
\emph{tau-leaping} method (discussed by Gillespie himself in
\cite{Gillespie01}) that consists of considering propensities
as constant for a certain (relatively large) 
interval of time $\tau$, and updating the 
state vector by a normal random variable only after that time $\tau$. 
In the pseudo-code below, we fixed a number of
firings $L$ instead of fixing a time interval $\tau$, but the
difference is not essential.
  We shall not detail the simulation code for normal distributions,
which is routine. We call \textsf{Normal}$(E,V)$ a function
simulating normally distributed random variables with vector expectation $E$
and covariance matrix $V$. 
\begin{algo}
Initialize\\ 
\tab Time scale $T\,\longleftarrow\,0$\\
\tab State vector $X\,\longleftarrow\,X(0)$\\
\tab Iterations number $L$\\
Repeat\\
\tab Compute propensities $\lambda_j(X)$ for $j=1,\ldots,m$\\
\tab Sum them to get $\lambda_{\bullet}(X)$\\
\tab Compute expectation and variance of time increment\\
\tab\tab $E_{time}\,\longleftarrow\,E_{time}+L/\lambda_{\bullet}(X)$\\
\tab\tab $V_{time}\,\longleftarrow\,V_{time}+L/\lambda_{\bullet}(X)^2$\\
\tab Compute probabilities: 
$p_j=\displaystyle{\frac{\lambda_{j}(X(s))}{\lambda_{\bullet}(X(s))}}$\\
\tab Compute expectation and covariance of state vector increments\\
\tab\tab $E_{state}\,\longleftarrow\, \sum_jp_j(\nu^+_j-\nu^-_j)$\\
\tab\tab
$V_{state}\,\longleftarrow\,\sum_jp_j(\nu^+_j-\nu^-_j-E_{state})
(\nu^+_j-\nu^-_j-E_{state})^t$\\
\tab\tab Multiply by loop length\\
\tab\tab\tab $E_{state}\,\longleftarrow\,L\times E_{state}$\\
\tab\tab\tab $V_{state}\,\longleftarrow\,L\times V_{state}$\\
\tab Update time scale: $T\,\longleftarrow\,
T+\mbox{\textsf{Normal}}(E_{time},V_{time})$\\
\tab Update state vector:
$X \longleftarrow X+
\mbox{\textsf{Normal}}(E_{state},V_{state})$\\
Until end of simulation
\end{algo}
As pointed out by Higham \cite{Higham08} the tau-leaping
method matches the Euler-Maruyama scheme
for the Chemical Langevin Equation:
\begin{equation*}
 \tag{CLE}
\begin{array}{lcl}
\dd C(t) &=&  \displaystyle{
\left(\sum_j \widetilde{\lambda}_j(C(s))
(\nu^+_j-\nu^-_j)\right)\,\dd t}\\[1.5ex]
&&\displaystyle{+
\frac{1}{\sqrt{N}} 
\left(\sum_j \sqrt{\widetilde{\lambda}_j(C(s))}(\nu^+_j-\nu^-_j)\,\dd W_j
\,\right)\;.}
\end{array}
\end{equation*}
It is a well know fact that the Euler-Maruyama scheme, like the Euler
scheme for ODE's, tends to be numerically unstable. At the expense of
a slight increase in computer time, one gets a  better
precision by using instead the Heun scheme. The book by Kloeden and
Platen \cite{KloedenPlaten97} is the indispensable reference on
numerical schemes for SDE's.

\vskip 2mm
Improvements on the Stochastic Simulation algorithms have generated
many works, and Gillespie
and his co-authors have been very active: Liu \emph{et al.}
\cite{Lietal08} give a review of existing methods; Cao \emph{et al.}
\cite{Caoetal04}
investigate numerical stability of leaping methods; Gillespie \emph{et
  al.} \cite{Gillespieetal05b,Gillespieetal05a} address the problem of
accelerating the simulation when propensities have very different
orders of magnitudes. In \cite{Gillespieetal09,Wuetal11}, the approximations to
the Michaelis-Menten equations are discussed, and the gain in computer
time of leaping is evaluated. The same problem of multiple time scales has also
been adressed by Ball \emph{et al.} \cite{Balletal06}, Haseltine and
Rawlings \cite{HaseltineRawlings02}, Liu and Vanden-Eijden
\cite{LiuVanden05}, McColluma \emph{et al.} \cite{McCollumaetal05}.
Other improvements include Bruck and Gibson's \cite{BruckGibson00}
study on simulation of large systems, and
the works by Ramaswamy and his co-authors 
\cite{Ramaswamy09,Ramaswamy10,Ramaswamy11a} on partial propensity.
Elf and Ehrenberger \cite{ElfEhrenberg03} replace simulation by an
evaluation of fluctuations around the solution of the RRE using
linear noise approximations.  Zhou \emph{et al.} \cite{Zhouetal08} 
have applied a
stochastic algorithm to coupled reactions with delays.
Recently, Zeron and Santillan
\cite{ZeronSantillan11} published a numerical study of a gene
network with negative feedback regulation, including the stability of
steady state.
\section{Computer integration}
The  ``Virtual Cell'' of Gunawardena \emph{et
  al.} in Harvard or de Jong \cite{deJong09} in Grenoble, is still a
beautiful dream; yet significant steps toward its realization have been made.

Regarding biochemical pathways, the process of knowledge accumulation
and sharing has long been very lively.
The ``Pathway Resource List'', or ``PathGuide'' \cite{Baderetal06} 
currently lists more than 400 databases on biological
pathways and molecular interaction, most of them
freely accessible. Examples include 
BiGG \cite{Schellenbergeretal10}
MetaCyc and BioCyc
\cite{Caspietal10},
KEGG \cite{Kanehishaetal04}, TRANSPATH \cite{Krulletal06},
BioCyc \cite{Krummenacker05},
Reactome \cite{Matthewsetal09}, PANTHER \cite{Mietal05}, PID
\cite{Schaeferetal09}. 
Several databases now come with integrated environments that offer
different vizualisations, like Cytoscape \cite{Shannon03}, or even
simulate the archived models, like BioModels \cite{Biomodels10} or 
Expasy \cite{Gasteigeretal03}. Kitano \cite{Kitano02} or Huang
\emph{et al.} \cite{Huangetal09} review
computational tools available in Systems Biology. 

In the accumulation process, the standardization issue was raised
very early. In 2000, Bader and Hogue
\cite{BaderHogue00} had already proposed BIND, a data specification
adapted to pathways.
In 2005, Cary \emph{et al.} \cite{Caryetal05} listed
170 existing databases and called for standard exchange formats to
successfully integrate data on a large scale. These were already under
development, and they integrate standard programming tools for the
simulation and treatment of gene networks and pathways: 
BioPAX \cite{Biopax10} SBML \cite{Huckaetal04}, CellML \cite{Milleretal10},
Biolingua \cite{Massar05}. Other programming toolboxes include COBRA
\cite{Beckeretal07}, GNA \cite{deJongetal03}, PATIKA \cite{Demiretal02},
and DIZZY \cite{Ramsey05}, the latest being more focused 
on stochastic simulation.   

Together with software realizations, informatics theoretical
researches attempted a formalization of languages and investigation
procedures. Fisher and Henzinger \cite{FisherHenzinger07} call
``Executable Biology'' that research area constructing 
computational models of biological systems. Kitano \cite{Kitano03}
proposes a graphical notation for biochemical networks. Danos and Laneve
\cite{DanosLaneve04} invent an abstract language for formal proteins.
Berthoumieux \emph{et al.}
\cite{Berthoumieuxetal11} study the identification of network models
though incomplete high throughput datasets.
For testing and experimenting purposes, Lok and Brent 
\cite{LokBrent05} construct an automatic generation tool 
for cellular reaction networks. 
Monteiro \emph{et al.} 
\cite{Monteiroetal09} propose a service oriented architecture for
integrated treatment of networks.
Schultz et al.
\cite{Schulzetaletal11} discuss the clustering of computational models based on
semantic annotations. Yamamoto \emph{et al.}
\cite{Yamamotoetal10} have developped an Artificial Intelligence
system, SOLAR, that includes reasonning tools for biological inference
on pathways.
\bibliographystyle{plain}
\bibliography{/home/ycart/recherche/Fournie/JJ}
\end{document}